\documentstyle[epsf,twoside,fleqn,espcrc2]{article}

\newcommand{\beq}{\begin{equation}}
\newcommand{\eeq}{\end{equation}}

\newcommand{\UKcol}{{\sf UKQCD collaboration}}
\def\Ord{{\cal O}}

\title{Determination of constant lattice spacing trajectories in lattice QCD}

\author{Alan C. Irving\address{Theoretical Physics Division,
         Department of Mathematical Sciences,
         University of Liverpool,
         PO Box 147, Liverpool L69 3BX, UK.},
	{\em UKQCD Collaboration}\thanks{supported by HPCI/EPSRC grant
	GR/K41663 and PPARC grant GR/K55745}
}

\begin{document}

\begin{abstract}
We argue that lattice simulations of full QCD with varying quark
mass are best conducted at fixed lattice
spacing rather than at fixed $\beta$.
We present techniques which
enable this to be carried out effectively, namely the tuning in bare
parameter space and efficient stochastic estimation of the fermion
determinant. Results and tests of the method are presented.
We discuss other applications of such techniques.
\end{abstract}

\maketitle

\section{CONSTANT LATTICE SPACING}
\label{sec:clatspac}
Hadron spectrum measurements by the \UKcol{}~\cite{MTlat97,UKQCDpaper} using
an improved action have shown a surprisingly strong dependence of
the effective lattice spacing on the bare quark mass.
Recent preliminary measurements of the Sommer scale $r_0$
using the non-perturbatively improved value of 
$c_{\rm sw}$ in $16^3\times 32$ simulations~\cite{MTlat98}
have confirmed this. See Table~\ref{tb:latspac}.
\begin{table}[h]
\caption{Lattice spacing determined from $r_0$~\cite{Sommer}
at $\beta=5.2$ for $\Ord (a)$-improved Wilson fermions}
\label{tb:latspac}
\centering
\begin{tabular}{lllll}
\hline
$\kappa$ &$.1330$ &$.1340$ &$.1345$ &$.1350$\\
$a$ [fm] &$.154(2)$ &$.140(4)$ &$.130(2)$ &$.1173(16)$ \\
\hline
\end{tabular}
\end{table}
Since the parameters in chiral perturbation theory inherit this
underlying $a$-dependence, 
this  complicates 
chiral extrapolations of simulation measurements and obscures
comparisons with quenched calculations. Indeed, one would not attempt
to compare quenched and full QCD at the same value of $\beta$. Because of
the strong and non-trivial dependence on the the bare quark mass,
one should avoid using {\it fixed} $\beta$ within the dynamical sector also.

With this in mind, we investigate how one might control the effective
lattice spacing by tuning the bare action parameters while the effects
of decreasing quark mass are studied. 
In previous
work~\cite{ACIJCS,ACIlat97} we have described how stochastic estimates
of the fermion determinant may be used to tune action parameters
so as to maintain fixed the value of some observable $F$. 
We would suggest that, in order to hold the effective lattice spacing
fixed, a suitable observable is the Sommer scale $r_0$. This is free
from complications due to valence quarks (partial quenching) and
chiral symmetry constraints.

For small
parameter changes, one may apply a first order cumulant 
expansion~\cite{ACIJCS}
\begin{eqnarray}  
<F>_2 = <F>_1 +<\tilde{F}\tilde{\Delta}_{12}>_1 + \dots\nonumber\\
\hbox{and}\quad\Delta_{12} \equiv S_1-S_2,\quad
\tilde{F}\equiv F - <F>\, \hbox{etc.}
\label{eq:F2F1}
\end{eqnarray}
Here, $<..>_i$ means the expectation value with respect
to action $S_i$. For the improved Wilson action,
\beq
\Delta_{12} \equiv 
(\beta_2-\beta_1)W_{\Box}+
T(\beta_2,\kappa_2)-T(\beta_1,\kappa_1)\, .
\eeq
The operator $T$ ($\equiv{\rm TrLn}M^{\dagger}M$) is $\beta$-dependent only 
when $c_{\rm sw}\neq 0$.
We may use ~(\ref{eq:F2F1}) to estimate
\begin{eqnarray}
\label{eq:dpdk}
{{\partial<\tilde{F}>}\over{\partial\kappa}}&\approx&
{<\tilde{F}{{\partial\tilde{T}}\over{\partial\kappa}}>}\qquad{\rm and}\\
{{\partial<\tilde{F}>}\over{\partial\beta}}&\approx& 
<\tilde{F}(\tilde{W}_{\Box}+
{{\partial\tilde{T}}\over{\partial c_{\rm sw}}}\dot{c}_{\rm sw})>\, .
\label{eq:dpdb}
\end{eqnarray}
Matching $F$ at
nearby points ($\beta_1,\kappa_1$) and ($\beta_2,\kappa_2$)
thus means solving 
\beq
{{\delta\beta}\over{\delta\kappa}}=
-{{<\tilde{F}\delta\tilde{T}>_1}\over
{\delta\kappa}<\tilde{F}\tilde{W}_{\Box}>_1}\, ,
\quad \delta\beta=\beta_2-\beta_1\,  {\rm etc.}
\label{eq:dbetdkap}
\eeq
(in the case where $c_{\rm sw}=0$). This in turn requires reliable (stochastic) estimates of 
$T$.

\section{LANCZOS QUADRATURE}
\label{sec:lancquad}
Bai, Fahey and Golub~\cite{BFG} have shown how to estimate the trace
of a matrix function via Gaussian quadrature using an elegant
relationship between orthogonal polynomials associated with
the measure and the Lanczos recursion scheme. We construct a noisy
estimator for TrLn$(M^{\dagger}M)$:
\beq
\hat{E}_T={1\over{N_{\phi}}}\sum_{i=1}^{N_\phi} I(\phi_i)\, ,
\quad I(\phi_i) = \sum_{j=1}^N\omega_j^2\hbox{Ln}(\lambda_j)\, .
\qquad 
\label{eq:EhatT}
\eeq
Here $\{\lambda_j^i\}$ $(j=1,2\dots N)$ are the eigenvalues of the
$N$-dimensional tridiagonal Lanczos matrix formed using the 
noise vector $\phi_i$ as a starting vector.
The weights $\{\omega_j^2\}$ are related to the corresponding
eigenvectors~\cite{BFG}. In fact, $\omega_j$ is just the
first component of the $j$th eigenvector of the tridiagonal
matrix. Rather than use bounds
for the truncation error associated with finite values of 
$N$~\cite{BFG}, we take advantage of the rapid (exponential)
convergence of the quadrature to obtain 6 figure accuracy for each noise
term $I(\phi_i)$ using typically $N=60$ to $100$ Lanczos iterations. This is
made possible by a remarkable stability to the round-off error and loss of
orthogonality normally associated with a standard Lanczos analysis of the
eigenvalue spectrum. We have noted an \lq exclusion principle\rq{}
associated with the quadrature weights $\omega_i$ which ensures that
only genuine eigenvalues of the tridiagonal system contribute to the
quadrature even in the presence of \lq ghost\rq{} 
and \lq spurious\rq{} eigenvalues~\cite{CSI}. 

\section{QCD MATCHING RESULTS}
\label{sec:QCDres}
Given an initial QCD simulation at $(\beta_1,\kappa_1)$ there are two
obvious ways to locate a nearby point $(\beta_2,\kappa_2)$ on the
constant $F$ curve: 
(A) via (\ref{eq:dbetdkap}) and Lanczos quadrature or 
(B) using brute force i.e. two further exploratory simulations at
shifted $\beta$ and shifted $\kappa$. It is straightforward to
estimate the relative work involved. We find that, to achieve the same
absolute error, method (A)
requires less than 5\%{} of the work for (B)~\cite{ISC}.
 
\begin{figure}[tb]
\begin{center}
\leavevmode
\epsfxsize=8.2cm
\epsfbox{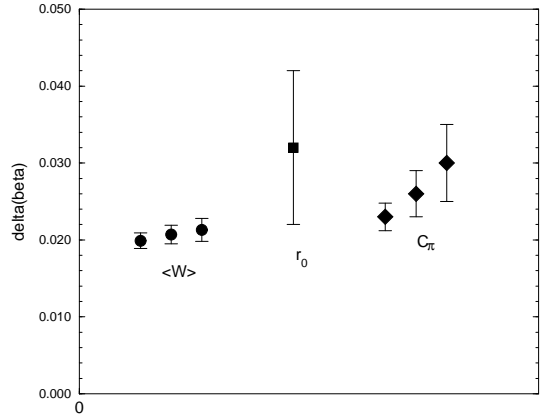}
\end{center}
\vspace{-1.5cm}
\caption{Predictions for the shift $\delta\beta$ from
reference point $(5.2,0.1340)$ required to keep
the specified observables constant when $\kappa$
is changed by $\delta\kappa=-0.0005$.
Circles correspond to Wilson loops ($1\times 1$,
$1\times 2$, $2\times 4$);
diamonds to $C_{\pi}(t)$ ($t=0$, $1$, $2$);
square to $r_0$.}
\label{fg:dbetsum}
\vspace{-0.5cm}
\end{figure}

We have conducted tests of the formalism starting with an ensemble
of 40 configurations of a $8^3\times 24$ lattice at 
$(\beta_1,\kappa_1)=(5.2,0.1340)$ generated via Hybrid Monte Carlo.
In particular, we have estimated the 
shift $\delta\beta_F$ required to compensate for a change
$\delta\kappa=-0.0005$ while the value of $F$ is held fixed.
Figure~\ref{fg:dbetsum} shows a compilation of shifts $\delta\beta_F$
for a variety of observables $F$: Wilson loops, the Sommer scale 
$r_0$~\cite{Sommer} and pion correlators $C_{\pi}(t)$.
In Table~\ref{tb:plaqtest} we show that the shift prediction
for the average plaquette ($1\times 1$ Wilson loop) is accurate.
An independent HMC simulation has been carried out at the
predicted matched point $(5.220,0.1335)$, using the appropriate value of
$c_{\rm sw}$. There is good agreement between the matched values of
$<P>$.
\begin{table}
\caption{Test of matching for the average plaquette $<P>$.
The additional error in the matched value is systematic, reflecting the
uncertainty in the matched beta value: $\delta\beta=0.0199(10)$.}
\centering
\begin{tabular}{l|l}
\hline
 $(\beta,c_{\rm sw },\kappa)$    &$<P>$\\
\hline
$(5.200,2.0171,0.1340)$ &$0.5286(3)$\\
$(5.220,1.9936,0.1335)$ &$0.5290(3)(4)$\\
\hline
\end{tabular}
\label{tb:plaqtest}
\end{table}
The partial derivatives (\ref{eq:dpdk},\ref{eq:dpdb}) 
of the average plaquette measured at
the reference point are 
\[
{{\partial<\tilde{P}>}\over{\partial\kappa}}=7.5(17)\, ,\qquad
{{\partial<\tilde{P}>}\over{\partial\beta}}=0.182(40)
\]

\begin{figure}   
\begin{center}
\leavevmode
\epsfxsize=8.2cm
\epsfbox{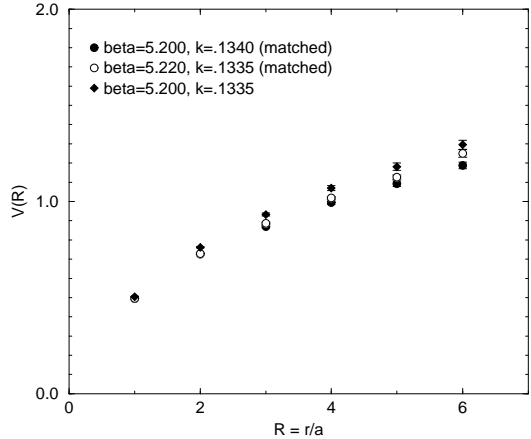}
\end{center}
\vspace{-1.5cm}
\caption{Static potential on an $8^3\times 24$ lattice
at $(5.2,0.1340)$ and $(5.220,0.1335)$ where
the average plaquette values match.}
\label{fg:potmatch}
\vspace{-0.5cm}
\end{figure}

In Figure~\ref{fg:dbetsum}, we see some evidence that the shifts
required to match observables which are sensitive to longer range physics are
somewhat larger (perhaps 50\%{}). With the modest number of
test configurations used, the statistical evidence is not strong.
However, Figure~\ref{fg:potmatch} confirms this point. 
One sees that matching the plaquette
ensures matching of the short distance potential but {\em not} the slopes
of the potentials, and hence $r_0$. Also shown in the figure is the
potential with a shift in $\kappa$ but not $\beta$. There is no matching
at any distance in that case.

\section{OTHER APPLICATIONS}
\label{sec:other}
The Lanczos quadrature technique allows one to build relatively
efficient stochastic estimators~(\ref{eq:EhatT}) of the fermion
determinant suitable for a variety of purposes including:
\begin{itemize}
\item matching observables (see above) 
\item tuning approximate actions~\cite{ACIJCS}
\item tuning exact algorithms based on these
\item choosing parallel tempering ensembles~\cite{BJoo}.
\end{itemize}
In particular, a suitably truncated version may be used to
model the low eigenmodes of the determinant~\cite{Duncan} while using other  
gauge invariant operators (e.g. Wilson loops)~\cite{ACIJCS} to account
for the remainder. Preliminary tests~\cite{ISC} have demonstrated that
relatively few Lanczos iterations are required to obtain a good
quantitative account of the fermion determinant. Both the
truncation parameters and the gauge loop parameters must be tuned,
using the above techniques, to obtain a useful representation of the
true QCD action and to ensure a good acceptance of configurations
proposed by the gauge part of the update. Such approximate actions can
of course be made exact via an accept/reject step, given sufficiently accurate
tuning~\cite{ACIJCS}.

\end{document}